\documentclass[prl,showpacs]{revtex4}
\usepackage{amsmath}

\usepackage{color}
\usepackage{epsf}
\usepackage{epsfig}
\usepackage{epstopdf}
\usepackage{graphicx}
\usepackage{latexsym}
\usepackage{bm}
\usepackage{amssymb}

\usepackage{epstopdf}

\usepackage{multirow}

\begin{document}

\begin{flushleft}
{\em Phys. Rev. Lett. 111, 233004 (2013)}
\newline
\end{flushleft}

\title{Vibrationally resolved decay width of Interatomic Coulombic Decay in HeNe}

\author{
F. Trinter$^1$,
J. B. Williams$^1$,
M. Weller$^1$,
M. Waitz$^1$,
M. Pitzer$^1$,
J. Voigtsberger$^1$,
C. Schober$^1$,
G. Kastirke$^1$,
C. M\"uller$^1$,
C. Goihl$^1$,
P. Burzynski$^1$,
F. Wiegandt$^1$,
R. Wallauer$^1$, 
A. Kalinin$^1$,
L. Ph. H. Schmidt$^1$,
M. S. Sch\"offler$^1$,
Y.-C. Chiang$^2$,
K. Gokhberg$^2$, \\
T. Jahnke$^1$
and R. D\"orner$^1$}
\email{doerner@atom.uni-frankfurt.de}

\affiliation{
$^1$ Institut f\"ur Kernphysik, J.~W.~Goethe-Universit\"at, Max-von-Laue-Str.1, D-60438 Frankfurt am Main, Germany \\
$^2$ Theoretische Chemie, Universit\"at Heidelberg, Im Neuenheimer Feld 229, D-69120 Heidelberg, Germany}
\maketitle

{\bf
We investigate the ionization of HeNe from
below the He 1s3p excitation to the He ionization threshold. We observe HeNe$^+$ ions with an
enhancement by more than a factor of 60 when
the He side couples resonantly to the radiation
field. These ions are an experimental proof of a
two-center resonant photoionization mechanism
predicted by Najjari et al. [Phys. Rev. Lett.
105, 153002 (2010)]. Furthermore, our data
provide electronic and vibrational state resolved
decay widths of interatomic Coulombic decay (ICD) in HeNe dimers. We find that the ICD lifetime
strongly increases with increasing vibrational
state.
}\\

In radio technology antennas are used to
efficiently collect energy from the
electromagnetic field. In the optical regime
nanoscale antennas have been developed 
\cite{Novtiny2012} and in
nature specialized antenna molecules efficiently
collect visible light to power biochemical reactions.
Schematically all these systems consist of the
antenna itself which couples to the radiation
field, a receiver which uses the energy
and a route to transport the energy between them.


The linear coupling of light to matter at
energies above the ionization threshold shows a
smooth energy dependence with stepwise increases
whenever a new shell opens. On top of this smooth
dependence of the ionization cross section are
different types of resonance features. For atoms
most notably there are e.g. Fano resonances created by doubly
excited states embedded in the continuum
\cite{Fano61prl}. In 2010 Najjari et al.
\cite{Najjari2010prl} predicted that for such a
resonant enhancement of the photoionization cross
section to occur, the resonant excitation state does not
need to be on the same subsystem which finally is
ionized. Instead one atom can act as an antenna which 
resonantly absorbs the light while a second atom can be the receiver
which finally gets ionized. This proposed antenna mechanism for the
interaction with the radiation field builds on
pioneering work by Gokhberg et al.
\cite{Gokhberg2005epl}, who showed that
excitation energy can be transferred from a
neutral atom to a neighbor in an interatomic Columbic decay (ICD)
\cite{cederbaum97prl,Jahnke04icdprl,marburger2003prl,Hergenhahn2011jesp}
like process. In chemistry intra-cluster energy transfer has been studied
since the mid 1980s under different names such as intermolecular autoionization \cite{Ukai89cpl}
or intramolecular penning ionization \cite{Kamke1985cpl}. It was also shown that it is feasible
to extend this physical scenario of two-center resonant photoionization to molecules \cite{golan2012jpcl}. 
The important question of the time scale of the energy transfer, however, could not be answered until today.

In the present experiment we demonstrate the smallest possible implementation of an antenna-receiver complex
which consists of a single (helium) atom acting as the antenna and a second (neon) atom acting as a receiver.
In this most simple antenna-receiver setup the antenna 
atom enhances the coupling to the radiation
field by more than a factor of 60. After being collected by the antenna the
energy is transferred to the receiver via ICD.
The fact that the studied system consists of just two atoms allows us to
investigate the physics of the energy transfer in high detail, i.e. on the level of a single vibrational
quantum state and to measure the energy transfer times. We find that the decay time strongly depends on the vibrational level which demonstrates the predicted strong dependence on internuclear distance.

\begin {figure}[t]
  \begin{center}
    \includegraphics[width=1.0\linewidth]{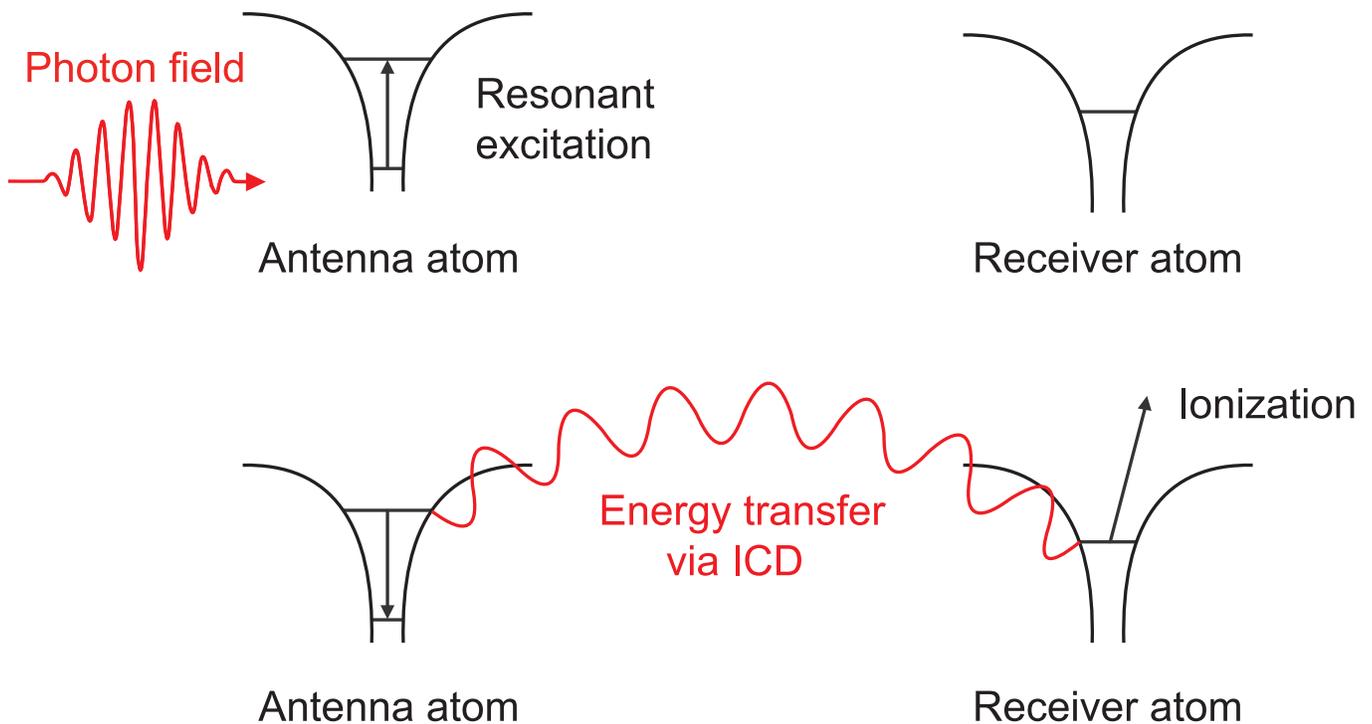}
    \caption{Schematic of the antenna mechanism. One atom resonantly absorbs a photon from the field storing its energy in as an electronic excitation.
    Before the state decays by fluorescence the energy is transferred
    by interatomic Coulombic decay to a neighboring receiver atom with an ionization potential
    lower than the resonance energy of the antenna atom where it leads to emission of an electron.
    Here we use He as an antenna atom and Ne as the receiver.
    }
  \label{fig1}
  \end{center}
\end {figure}

For this antenna mechanism to be unambiguously identified in an experiment the two atoms
involved must
have significantly different ionization
thresholds. The antenna system has to possess an
excited bound state with a large oscillator
strength which is energetically located above the
ionization threshold of the receiver system, as
illustrated in Figure 1. In addition the energy
transfer mechanism has to be fast compared to the
radiative relaxation of the antenna. In the
present work we choose the HeNe dimer in the gas
phase as a showcase system which has the required
energetic properties and we measure the time and the
efficiency of the energy transfer.

In HeNe the antenna (He) and receiver (Ne) are
weakly bound by the van der Waals force at an
equilibrium distance of 3.0 \AA. The binding
energy is only 2 meV which is tiny compared to an ionization
potential of 21.564 eV of Ne and 24.587 eV of He.
The singly excited state He(1s3p)$^1P^0$ at
23.087 eV is the first dipole allowed He state
above the Ne ionization threshold, i.e. the first state above the threshold that is
quantum-mechanically allowed to couple to the photon field. Thus, 
HeNe is a good candidate to experimentally
search for the predicted antenna effect.

The experiments have been performed at beamline UE112-PGM-1 at
the synchrotron radiation facility BESSY
(Berlin) in single bunch operation using the
COLTRIMS technique \cite{Ullrich03rpp}. The HeNe
dimers were produced by coexpanding a $80/20~\%$
mixture of He and Ne gas through a cooled nozzle
at 52 K at a driving pressure of 5 bar. This
resulted in a fraction of about $3.2~\%$ HeNe dimers
and $4.4~\%$ Ne$_2$ dimers. The supersonic gas jet
passed two skimmers (0.3 mm diameter) and was
crossed with the photon beam at right angles.
Ions and electrons created in the interaction
region were guided by a weak electric (7.5 V/cm)
and a parallel magnetic (7.8 Gauss) field onto
two position sensitive detectors with delayline
readout \cite{Jagutzki02nim}. The time-of-flight
and position of impact of all particles were
measured. The time-of-flight spectrum of the measured ions
(Figure 2) shows a clear separation of $^{20,22}$Ne$^+$ ions,
He$^{20}$Ne$^+$ ions and He$^{22}$Ne$^+$ ions. In the present experiment the electron
detection and the information on the position of impact of the ions on the ion
detector was solely used for background
suppression. The photon beam energy was scanned
using the monochromator at an energy resolution which was
measured in situ during the scans to be 1.7 meV
(FWHM) by monitoring the strong signal of
fluorescence photons from the He(1s3p)$^1P^0$
resonance of the $80~\%$ of Helium atoms in our gas
jet (see Figure 3).

The strength of the antenna effect can most
directly be seen in Figure 2. Off the He
resonance the ratio of Ne$^+$ counts to HeNe$^+$
counts is about $3.2~\%$, which is the
HeNe content in our gas jet. On the resonance
(Figure 2b), however, HeNe$^+$ is the highest peak.
This enhancement of the ionization of HeNe as
compared to pure Ne far below the He ionization
threshold is caused by the He atom acting as an
antenna coupling efficiently to the light field.
The counts in the area labeled 'dissociation' are
Ne$^+$ ions with kinetic energies of a few 100 meV which are visible as a halo to the Ne$^+$ line.
They show the same resonances as the HeNe$^+$
signal. We attribute them to dissociation of
HeNe$^+$ after the electron emission.

\begin {figure}[t]
  \begin{center}
    \includegraphics[width=1.0\linewidth]{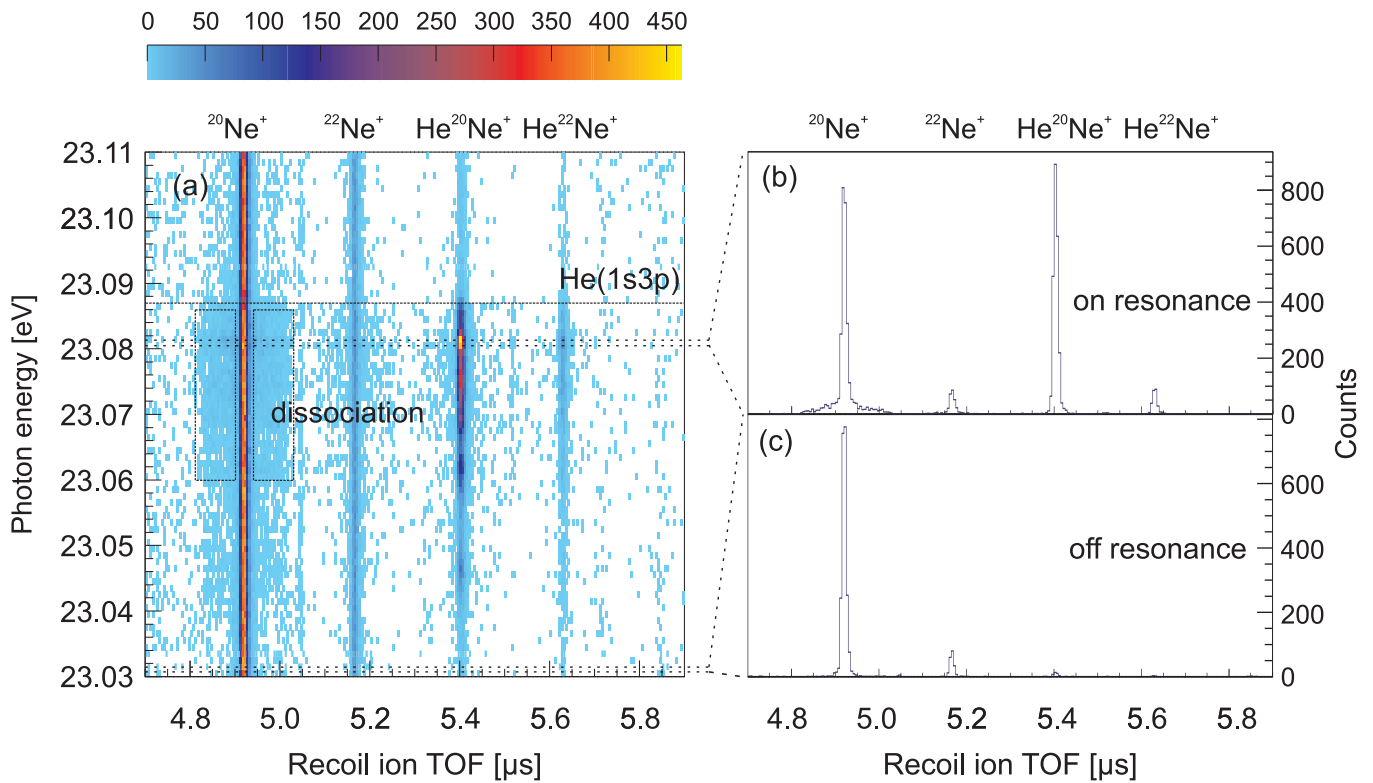}
    \caption{Photoionization of a supersonic jet containing He ($80~\%$) and Ne ($20~\%$)
    and about  $3.2~\%$ HeNe dimers. (a) Ion time-of-flight versus photon energy in the vicinity of the He(1s3p) resonance.
    (b) Ion time-of-flight for the photon energy range 23.0805 - 23.0815 eV (on resonance).
    (c) Ion time-of-flight for the photon energy range 23.0305 - 23.0315 eV (off resonance).
    (b) and (c) are projections from the data shown in (a).}
  \label{fig2}
  \end{center}
\end {figure}

\begin {figure}[t]
  \begin{center}
    \includegraphics[width=1.0\linewidth]{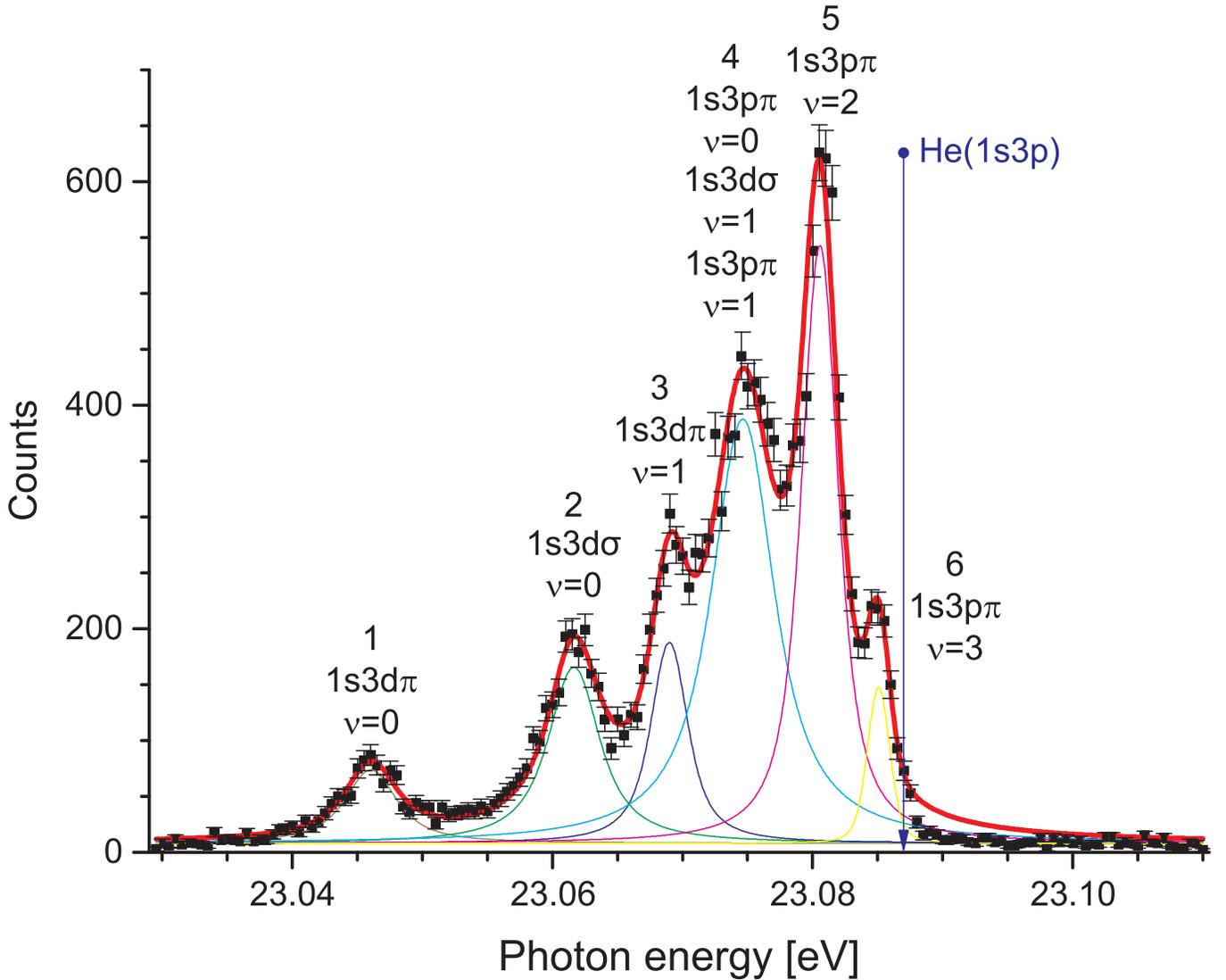}
    \caption{HeNe$^+$ photoion yield as function of photon energy in the vicinity of the He(1s3p) resonance. The fits have been made using a multiple Voigt fit (convolution of Lorentzian and Gaussian) with a fixed Gaussian width of 1.7 meV which is the measured
    energy resolution of the photon beam. The line indicates the energy of the He(1s3p) in an isolated He atom. For state assignment see text and Figure 4.}
  \label{fig3}
  \end{center}
\end {figure}

\begin {figure}[t]
  \begin{center}
    \includegraphics[width=1.0\linewidth]{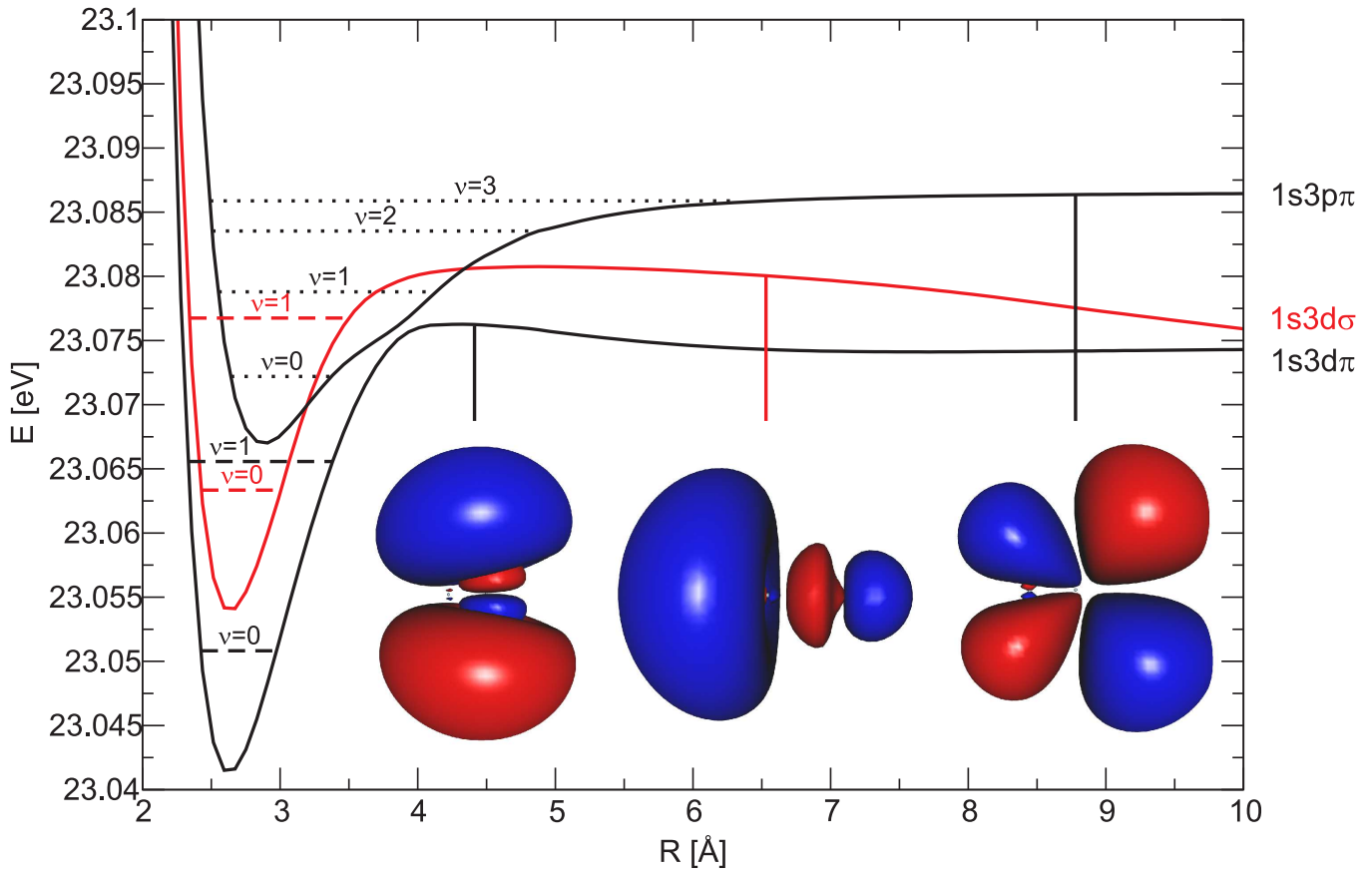}
    \caption{Potential energy curves of the excited states of HeNe and the natural orbitals of the excited electron at R=3.04 \AA \cite{orbitals}. The atom on the right is He. The 1s3p$\sigma$ state is off the scale, it lies about 100 meV above the states shown here.}
  \label{fig3b}
  \end{center}
\end {figure}

\begin {figure}[t]
  \begin{center}
    \includegraphics[width=1.0\linewidth]{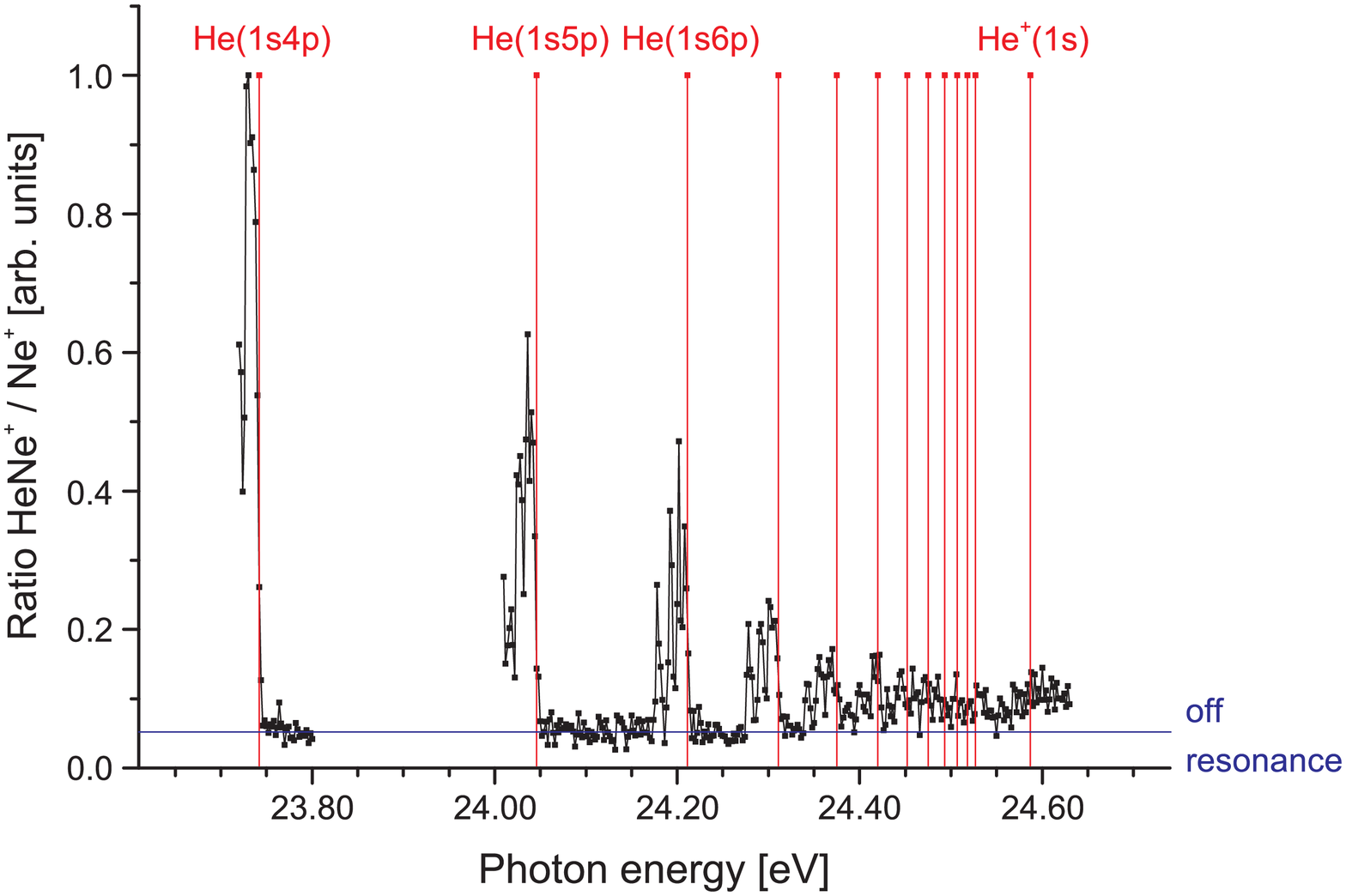}
    \caption{HeNe$^+$ photoion yield as function of photon energy from the He(1s4p) to the He ionization threshold. The blue baseline results from the non-resonant ionization of the Ne atom.}
  \label{fig4}
  \end{center}
\end {figure}

For a more detailed investigation of the antenna
effect we show the count rate of HeNe$^+$ as function of the photon
energy in the vicinity of the He(1s3p)$^1P^0$
resonance in Figure 3.  In contrast to
the flat energy dependence of the Ne$^+$ rate the
HeNe$^+$ shows a multiple peak structure slightly below the He(1s3p)$^1P^0$ resonance (see also Figure 5).

This energy shift is induced by the van der Waals
environment of the He atom.
In addition to the energy shift, the influence of the neighbor 
also mixes p and d states making states of mainly d character 
optically accessible.
To identify the peaks we calculated the potential energy surfaces and electronic transition moments using the EOM-CCSD(T) method \cite{Piecuch} and  EOM-CCSD method \cite{orbitals}, respectively, with v5z/4s5p5d (He), av5z (Ne) basis sets (Figure 4). The electronic decay widths were computed using the Fano-Stieltjes method \cite{Kopelke} with v5z/4s5p5d (He), v5z/4s4p4d (Ne) basis sets employed.
The energies and widths of the vibrational states were obtained by diagonalizing the resulting non-Hermitian nuclear Hamiltonian \cite{Sisourat_PRA}.
Table 1 shows the assignment of all of the experimental peaks based on 
comparison with this calculation.



The energy absorbed by the He resonance is
transferred to the neighboring Ne site from where a 2p
electron is emitted. The lifetime of the radiative decay of the He(1s3p)$^1P^0$ excitation resonance in
isolated He is 1.76 nsec corresponding to
a width of $2.35 \cdot 10^{-3}$ meV \cite{Johansson03}. The individual peaks in
Figure 3 are much broader, showing that the
resonantly excited state decays much faster in the vicinity
of the neon atom than in an isolated helium atom.
In order to obtain a lifetime for each of the
peaks we have fitted the spectrum with a sum of
six Lorentzians convoluted with a single Gaussian.
The width of the Gaussian given by the energy
resolution of the photon beam is the same for
each of the peaks. An intrinsic check for that width is given online by measuring 
the fluorescence light from the atomic He(1s3p) component in our jet.
Its peak position also provides an in situ
absolute calibration of our energy scale (see blue line in Figure 3).

The temporal behavior of ICD and ICD-like processes is of interest since the first predictions of ICD: The energy transfer time, i.e. the ICD width depends on the electronic state and on the internuclear distance of the participating atoms or molecules. Furthermore, the number of atomic neighbors that surround the excited particle determines the efficiency of ICD: In some cases, where that number is high, the ICD width becomes comparable to that of a typical Auger decay. In these cases ICD occurs within a few fsec \cite{Averbukh2006}. Pioneering work by \"Ohrwall et al. measured a lifetime of Ne$^+$(2s$^{-1}$) in the bulk of larger Ne clusters of 6 fsec \cite{ohrwall2004prl} confirming the time scale on which ICD takes place for the first time. If the excitation mechanism for a decaying state is energetically broadband or the decay width is broader than the vibrational spacing, a coherent vibrational wave packet will be formed upon excitation and the ICD occurs while the excited dimer shrinks or expands as shown in theoretical work \cite{Scheit, Sisourat_PRL, Sisourat_NatPhys}. Two recent experiments on ICD of rare gas dimers addressed this scenario. Schnorr et al. \cite{Schnorr13} measured a mean value of the decay time of excited neon dimers integrating over internuclear distances and approximating the non-exponential decay with an exponential decay function in a pump-probe experiment. The first time-resolved measurement of ICD was reported by Trinter et al. \cite{Trinter13prl}. They measured the time dependence of the internuclear distance and were able to show that in the coherent excitation scenario the decay function of ICD is as predicted \cite{Scheit} non-exponential. 
In contrast to the scenario described above, our present experiment in the energy domain uses narrow band excitation and resolves the electronic and the vibrational states. The measured vibrationally resolved widths are shown in Table 1. Peaks 4-6 correspond to the second, third and fourth vibrational level of 1s3p$\pi$ (peak 4 is a mixture of 1s3p$\pi$ $\nu$=0, 1s3p$\sigma$ $\nu$=1 and 1s3p$\pi$ $\nu$=1 states). It turns out that the lifetime increases from 160 to 1100 fsec with increasing vibrational level. This is a direct consequence of the increase of internuclear distance with increasing vibrational level. We find the shortest lifetime for the first vibrational state of 1s3d$\pi$. This might be surprising since in an isolated atom this state would even be dipole forbidden. In the dimer obviously the Stark mixing between p and d states and the small internuclear distance result in these very short decay times.
 
\begin{table}
    \centering
        \begin{tabular}{c|c|c|c|c|c|c|c|c} & \multicolumn{4}{c|}{Experiment} & \multicolumn{4}{|c}{Theory}														     \\
        																\hline
        																Peak & Energy  & Width               & Lifetime & Inten-    & Assignment   & Energy	& Width & Inten-			     \\
                                             & (eV)    & (meV)               & (fsec)   & sity      &           	 & (eV)		& (meV) & sity				     \\				
                                    		\hline
                                     		1    & 23.0460 & 5.0$\pm$1.5         & 130      &  5 \%     & 1s3d$\pi$    $\nu$=0 & 23.05082	& 3.74	& 7.7 \%   \\  \hline     						
                                    		2    & 23.0617 & 4.5$\pm$1.0         & 150      & 13 \%     & 1s3d$\sigma$ $\nu$=0	& 23.06334	& 5.68	& 7.2 \%       		       \\	\hline		
                                    		3    & 23.0690 & 3.0$\pm$1.0         & 220      & 10 \%     & 1s3d$\pi$ $\nu$=1   & 23.06557	& 2.90	& 17.4 \%                 \\ \hline

	\multirow{3}{*}{4}    & \multirow{3}{*}{23.0746} & \multirow{3}{*}{4.0$\pm$1.5}         & \multirow{3}{*}{160}      & \multirow{3}{*}{38 \%}     & 1s3p$\pi$    $\nu$=0 & 23.07220	& 1.41	& 0.9 \%   \\
	& & & & & 1s3d$\sigma$ $\nu$=1	& 23.07675	& 3.66	& 23.3 \% \\
	& & & & & 1s3p$\pi$ $\nu$=1	& 23.07879	& 0.85	& 4.4 \% \\ \hline					
                                     		5    & 23.0806 & 2.5$\pm$1.2         & 260      & 30 \%     & 1s3p$\pi$    $\nu$=2 & 23.08353	& 0.62	& 26.6 \%   \\   					\hline
                                     		6    & 23.0851 & 0.6$^{+1.5}_{-0.3}$ & 1100     &  4 \%     & 1s3p$\pi$    $\nu$=3 & 23.08588	& 0.22	& 12.6 \%   \\

        \end{tabular}
    \caption{Peak positions, energies, lifetimes and relative peak intensities for the six peaks in Figure 3 as well as assigned excited states of HeNe from theory including energies, widths and relative peak intensities.
    The experimental numbers have been obtained using a multi Lorentzian fit convoluted with the Gaussian width of our photon beam, which has
     been measured independently through the fluorescence light from the atomic He(1s3p) resonance. The errors of the widths are conservative estimates based on fitting with different seed values and different fitting software.}
    \label{tab:Table1}
\end{table}

Figure 5 displays an extended scan of the photon
energy from the He(1s4p) resonance to the He
single ionization threshold. Each of the dipole
allowed excited states shows the antenna effect
as the photon energy dependent ratio of HeNe$^+$ to Ne$^+$ reveals.

In conclusion we have experimentally shown that a
single atom can act as a highly efficient antenna
to absorb energy from a light field and pass
the energy to a neighboring receiver atom within
a few hundreds of femtoseconds. The
vibrationally resolved measurement of the
resonance width for a selected electronic state provides a benchmark for future
calculations of the underlying energy transfer
mechanism of ICD. Our findings yield (to our knowledge) the first vibrationally resolved lifetimes of ICD after narrow band excitation.

\textbf{Acknowledgements} We acknowledge the support of RoentDek Handels GmbH. The work was supported by
Research Unit 1789 of the Deutsche
Forschungsgemeinschaft. The experimentalists thank
Alexander Voitkiv and Carsten M\"uller for
proposing this experiment and providing a
theoretical estimate of the magnitude of the
expected effect. We thank Gregor Schiwietz and
the team at BESSY for operating the beamline and
providing the photon beam.


\bibliographystyle{unsrt}

\end{document}